\documentclass[a4paper,10pt]{article}
\usepackage[utf8]{inputenc}

\title{Application of Quantum Field Theory}
\author{Ashaq Hussain Sofi and Muhammad Ashraf Shah\\
Department of Physics, National Institute of Technology, \\ Srinagar, Kashmir-190006,  India}

\begin{document}

\maketitle

\begin{abstract}
In this paper we will analyse some interesting structures that occur in scalar quantum field theory. 
We will quantize this theory using path integrals. 
We will analyse the Bogomolny Bound for scalar quantum field theory in two dimensions. 
We will also analyse the generalization of this result to fractional powers of the differential operator. 
\end{abstract}

\section{Introduction}
Phase transition can be described by quantum field theory. This is because it is described by Gibbs free energy, which can be written 
as $G =  A(T) + B(T) M^2 + C(T) M^4...$, where $T$ is the temperature and $M$ is given by an integral over the spin variable. 
The Green's function for this system is given by 
\begin{equation}
 D(x) = \int d^3k \frac{H_0 e^{ikx}}{|k|^2 + 2b(T -T_c)},
\end{equation}
where $T_c$ is the temperature at which phase transition occurs. Thus, scalar 
quantum field theories have important applications 
in condensed matter physics. In fact, phase transitions can have important technological applications. For 
example in the formation of antiparticles. 
Transparent conducting oxides (TCOs) have attracted a great deal of attention in material research for the past few years
due to their numerous technological applications. The important and most commonly studied TCOs are the oxides of zinc, tin
and indium
to their remarkable combination of high transparency in the visible region and high conductivity \cite{1}-\cite{2}.
For example indium tin oxide (ITO) is a  
semiconductor and is widely used a coating material in energy saving lamps, solar cells,liquid crystal displays, automobile
instruments
due to their excellent electro-optical properties such as high electrical conductivity and transparency to light. It has
been investigated 
for applications as an anti-static coating material for spaceship \cite{3} and as a photo catalyst \cite{4}.
The performances of ITO material rely on the preparation of homogeneously mono dispersed ITO antiparticles.
These theories also interesting structures in (1+1)
dimensions and we will analyse few of them in this paper. 
The most fundamental quantity in any field theory is the classical action $S$
which is the spacetime integral of the Lagrangian density $\mathcal{L}$.
In field theory it is customary to call Lagrangian density $\mathcal{L}$ just
Lagrangian and we will adopt this usage. So the action $S$ is given by
 \begin{equation}
  S = \int d^2 x \mathcal{L}.
 \end{equation}
\begin{equation}
 \mathcal{L} = \frac{\sqrt{-g}}{2}(\nabla_a\phi \nabla^a\phi + m^2 \phi^2),
\end{equation}
where $m$ is the mass and $g$ the determinant of the metric.
This is the field theory we will analyse in this paper.

\section{ Scalar Field Theory }
Now if we assume that the field vanishes on boundaries, then we can show that the action can be written as follows:
 \begin{equation}
  S = \int d^2 d^2x' \frac{- \sqrt{-g(x)}\sqrt{-g(x')}}{2}\phi(x) E(x,x')\phi(x'),
 \end{equation}
 where
 \begin{equation}
  E(x,x') = ( \nabla^2 - m^2 ) \delta (x, x').
 \end{equation}
If $J (x)$ is the source of the field $\phi (x)$ then we define $J\phi$ as follows:
 \begin{equation}
  J\phi = \int d^4x \sqrt{-g}J(x) \phi(x).
 \end{equation}
We call $Z[J]$  the vacuum-to-vacuum  transition amplitude and it is given by
 \begin{equation}
  Z[J] = N\int D\phi \exp i [S + J\phi ].
 \end{equation}
 Here $N$ is a normalization constant, such that $Z[J]$ is normalized as follows:
 \begin{equation}
  Z[0] =1.
 \end{equation}
It can be shown that  $Z[J]$ is given by 
\begin{equation}
 Z[J] = \exp \frac{-i}{2} \int d^2x d^2x' \sqrt{-g(x)}\sqrt{-g(x')} J(x) G(x,x')J(x'),
\end{equation}
where $G(x,x')$ is  the inverse of $E(x, x')$,
\begin{equation}
 \int d^2 x' \sqrt{-g}  E (x, x') G(x',x'') =  \delta (x',x'').
\end{equation}
Then the two-point function  is given by
\begin{equation}
- \frac{\delta}{\delta J(x)}\frac{\delta}{\delta J(x')} Z[J]\left. \right|_{J= 0} = i G(x, x').
\end{equation}
Now we can define  a quantity called conjugate momentum current $\pi^c$ as follows:
 \begin{equation}
  \pi^c = \frac{1}{\sqrt{-g}}\frac{\partial \mathcal{L}}{\partial \nabla_c \phi},
 \end{equation}
and 
 \begin{equation}
  J^c = i [\phi_1^* \pi^c_2 - \phi_2 \pi^{*c}_1].
 \end{equation}
 Then we have \cite{ashr},
 \begin{eqnarray}
\nabla_c J^c   &=& 0.
 \end{eqnarray}
Thus the current $J^c$ is current (1+1) dimensional field theory as 
 \begin{equation}
  (\phi_1, \phi_2) = \int dx J^c_{(\phi_1, \phi_2)}.
 \end{equation}
 and so
\begin{equation}
 \frac{d}{dt}(\phi_1, \phi_2) = \int d^3 x \partial_0 (\sqrt{-g}J^0).
\end{equation}
Now as 
\begin{equation}
 \nabla_c J^c = \frac{1}{\sqrt{-g}} [\partial_0 (\sqrt{-g} J^0) + \partial_i (\sqrt{-g} J^i)] =0.
\end{equation}
So, we have 
\begin{equation}
\int d^4 x \partial_0 (\sqrt{-g}J^0) = 0.
\end{equation}
So this inner product does not vary with time.

Let $\phi_n$ and $\phi^*_n$ be a complete set of solutions to the field equations, 
then by definition we can expand $\phi$ as follows:
 \begin{equation}
  \phi = \sum_n [a_n \phi_n + a^*_n \phi^*_n ].
 \end{equation}
Here the sum is a shorthand notation and may contain integrals as well, for non-compact spacetime.
 Now the state $|0\rangle$ is  the state annihilated by $a_n$
 \begin{equation}
  a_n |0\rangle = 0.
 \end{equation}
 Now the two-point function is given by \cite{ashr}
 \begin{equation}
   G(x,x') = \sum_{nm} \psi_n \psi'_m M^{-1}_{nm}.
 \end{equation}

\section {Bogomolny Bound}
This is an interesting bound that exists in field theory.  Here we analyse the bound in details. 
The above Lagrangian can be modified to 
\begin{equation}
 \mathcal{L} = \frac{1}{2}
\partial_{\mu}\phi \partial^{\mu}\phi- {U}(\phi),
\end{equation}
this bound is attained if the field equations are static i.e., 
$\dot{\phi} = 0$ and 
\begin{equation}
 \phi' = \pm \sqrt{2U},
\end{equation}
where $\dot{\phi}$ is the time derivative of the field and $\phi'$ is the 
spatial derivative of the field.  The solutions with a positive 
sign are called kinks and those with negative sign are called anti-kinks. 
For these solutions the energy density from $\phi'/2$  and $U(\phi)$ are 
point wise the same. This is a stronger statement than viral theorem. 

The simplest example of a kink is probably a $\phi^4$-theory
for which the potential is given by 
\begin{equation}
{U(\phi)} = \lambda (m^{2} - \phi^{2})^{2}
\end{equation}
with $\lambda$ a real positive constant. 
The corresponding field equation for this model is
\begin{equation}
\partial_{\mu}\partial^{\mu}\phi - 4\lambda (m^{2} -\phi^{2})\phi = 0.
\end{equation}
There are two vacuum states for this model denoted by $\phi_{\pm}$, where 
$ \phi_+ = m $ and $\phi_- = -m$.
Here $\phi_{\pm}$ are the field values at $ x = \pm\infty $. 
Now we can define the topological charge 
as 
\begin{equation}
N = \frac{\phi_{+} -\phi_{-}}{2m}.
\end{equation}
We can write $N$ 
as the integral over space of a topological charge density 
\begin{equation}
N = \frac{1}{2m}\int_{-\infty}^{\infty}\phi' dx.
\end{equation}
Now when $N =0$, the field interpolates
  between the same vacuum, 
so it lies in the same topological sector as on 
of the vacuum solutions, $\phi(x)=\pm m$ as it can be continuously 
deformed to it. However, when  $N=1$,
 the minimal energy solution is the 
kink which interpolate between the two vacuum $\phi_{-}$ and $\phi_+$
 as $x$ increases
 from $-\infty$ to $\infty$. Furthermore, for the anti-kink $N =-1$ and it is obtained by 
making the replacement $\phi \to -\phi$. There are no multi-kink solutions for this 
model. Now, 
the Bogomolny energy bound for the $\phi^{4}$ is given by 
\begin{eqnarray}
E &\geq& \left| \int_{\phi-}^{\phi+}\sqrt{2\lambda}(m^{2} - \phi^{2})d\phi 
\right|
\nonumber \\ &=&  \left| \sqrt{2\lambda}[m^{2}\phi 
- \frac{1}{3}\phi^{3}]_{\phi-}^{\phi+}\right|\nonumber \\ 
              & =& \frac{4}{3}m^{3}\sqrt{2\lambda}| N|.\label{q}
\end{eqnarray}
For both the kink and anti-kink $|N| =1$ and 
so, the equality  is attained when the Bogomolny bound is 
saturated and in that case, we have 
\begin{equation}
\phi' = \sqrt{2\lambda}(m^{2} - \phi^{2}). 
\end{equation}
The solution to this  equation is given by 
\begin{equation}
\phi(x) = m \tanh (\sqrt{2\lambda}m(x-a)),
\end{equation}
where $a$ is an arbitrary constant of integration.  
So, now the energy density of the kink can be written as 
\begin{eqnarray}
\mathcal{E}&=& \frac{1}{2}\phi'^{2} 
+ \lambda(m^{2} -\phi^{2})^{2} = 
2\lambda m^{4} \rm{sech}^{4}(\sqrt{2\lambda}m(x-a))
\end{eqnarray}
From this we can calculate the energy of the Kink, and this is the same as 
one would expect  with  $|N| =1$,
\begin{equation}
E = \int_{-\infty}^{\infty}\mathcal{E}dx =
 \frac{4}{3}m^{3}\sqrt{2\lambda}.
\end{equation}

Now we can study this bound for the Sine-Gordon model.  
The Sine-Gordon model is a completely integrable 
field theory in (1+1) dimensions. It is known to have  multiple kinks solutions. 
For this theory the potential 
is given by 
\begin{equation}
U(\phi)=(1-\cos\phi).
\end{equation}
The equation of motion for this theory  can now be written as 
\begin{equation}
\partial_{\mu}\phi\partial^{\mu}\phi-\sin \phi=0.
\end{equation}
This theory has multi-vacuum states given by $\phi = 2\pi n$, where $n$ is any
integer. Now if $\phi_{\pm}$ be any two vacuum states in this theory, then 
the topological charge is given by 
\begin{equation}
N = \frac{\phi_{+} -\phi_{-}}{2\pi}.
\end{equation}
We can again write $N$ 
as the integral over space of a topological charge density 
\begin{equation}
N = \frac{1}{2m}\int_{-\infty}^{\infty}\phi' dx.
\end{equation}
Here $N$ counts the number of solicitous. 

The Bogomolny bound is now given by 
\begin{eqnarray}
E&\geq& \left| \int_{0}^{2\pi N}
\left[ 2\sin \frac{\phi}{2} \right] d\phi \right|\nonumber \\
   &=& 8|N|. \label{1}
\end{eqnarray}
This bound is attained when 
\begin{equation}
 \phi' = \pm 2 \sin \frac{\phi}{2}. 
\end{equation}
We can again calculate the energy density from this equation and then use that energy density to 
 calculate the energy of the kink.

\section{Generic Formalism for K-Field Models }
Many generalizations to Sine-Gordon system have been made. 
One of the interesting model that we will analyse here is the $K$-fields model \cite{f2a}-\cite{f4a}.
In these models the kinetic term is non-canonical and their solicitous have interesting applications. 
Their analytical solicitous have been recently investigated \cite{f4a}.  
The action for a generic $K$-field theory is given by 
\begin{equation}
S=\int d^2 x \mathcal{L}= \int d^2 x \left(F(X)-U(\phi)\right),
\end{equation}
where $X$ is the conventional kinetic term given by 
\begin{equation}
 X = \frac{1}{2} \partial^\mu \partial_\mu. 
\end{equation}
Here $F$ is any fractional power of this operator. 
Fractional differential operators are expressed as follows, 
\begin{equation}
 \partial^a f(x) = (\Gamma(a))^{-1} \int_0^x (x-t)^{a-1} F(t) dt
\end{equation}

Here the kinetic term $F(X)$ is an arbitrary function of the of this conventional kinetic term. 
The  momentum-energy tensor associated with this term is given by 
\begin{equation}
\mathcal{T}^{\mu\nonumber \\u}= F_X \partial^{\mu}\phi \partial^{\nonumber \\u} \phi -\eta^{\mu\nonumber \\u}\mathcal{L}.
\end{equation}
For a static configuration it is possible to define 
\begin{equation}
 F_X\equiv \frac{dF}{dX}.
\end{equation}
In this case the  non-vanishing components of the energy-momentum tensor  are given by  
\begin{eqnarray}
\mathcal{T}^{00}&=&\varepsilon=-\mathcal{L}=-F(X)+U(\phi), \nonumber \\
\mathcal{T}^{11}&=&-p=F(X)-2XF_X,
\end{eqnarray}
here $\varepsilon$ and $p$ are the energy density and the pressure respectively.
The BPS state is pressure-less source i.e, $ p =0$. 
As a concrete  example to this let us consider the following model 
\begin{equation}
L_{L}=X |X|-\frac{3}{4}\sin^{4}(\overline{\phi}_{0}/2)\cos^{4}(\overline{\phi}_{0}). 
\end{equation}
The model  has an analytical solution given by 
\begin{equation}
\phi_L(x)=2 \cos\left[\frac{\overline{\phi}_{0}(x)}{2}\right].
\end{equation}
Similarly we could consider the following model 
\begin{equation}
L_{W}=X |X|-\frac{3}{4}\cos^{4}(\overline{\phi}_{0}/2)\cos^{4}(\overline{\phi}_{0}),
\end{equation}
has an analytical solution given by
\begin{equation}
\phi_W(x)= 2 \sin\left[\frac{\overline{\phi}_{0}(x)}{2}\right]. 
\end{equation}
These two models can be merged into a single model. The  Lagrangian density for that single model can be
expressed as \cite{f4a}
\begin{equation}
L=X |X|-\frac{3}{4}\left(1-\frac{\phi^2}{4}\right)^2\left(-1+\frac{\phi^2}{2}\right)^4. 
\end{equation}
The BPS state for this model is given by 
\begin{equation}
-3X |X|=U.
\end{equation}
This can be expressed as 
\begin{equation}
  \phi^{'4}=\left(1-\frac{\phi^2}{4}\right)^2\left(-1+\frac{\phi^2}{2}\right)^4.
\end{equation}

\section{Conclusion}
In this paper we analysed the formation of ITO using quantum field theory. 
This field theory arises at the critical point for the second order phase transition. 
It may be noted that gauge symmetry can also arise in  a quantum field theory. 
It would be interesting to analyse this symmetry further. 
\cite{a}-\cite{wd1}. It would be interesting to see which results can be used and have physical 
implication.

\end{document}